\documentclass[11pt,twoside]{article}
\usepackage{asp2010}

\resetcounters

\begin{document}

\title{Probing magnetic fields with GALFACTS}
\author{Samuel J. George$^{1,2}$, Jeroen M. Stil$^{2}$, Mircea Andrecut$^{2}$ , A. Russ Taylor$^{2}$
\affil{$^1$Astrophysics Group, The Cavendish Laboratory, JJ Thomson Avenue, University of Cambridge,
Cambridge, CB3 0HE, UK}
\affil{$^2$Institute for Space Imaging Science,\\ University of Calgary, Alberta, T2N 1N4, Canada}}

\begin{abstract}
GALFACTS is a large-area spectro-polarimetric survey on the Arecibo Radio telescope. It uses the seven-beam focal plane feed array receiver system (ALFA) to carry out an imaging survey project of the 12,700 square degrees of sky visible from Arecibo at 1.4 GHz with 8192 spectral channels over a bandwidth of 300 MHz sampled at 1 millisecond. The aggregate data rate is 875 MB/s. GALFACTS observations will create full-Stokes image cubes at an angular resolution of 3.5' with a band-averaged sensitivity of 90 $\mu$Jy, allowing sensitive imaging of polarized radiation and Faraday Rotation Measure from both diffuse emission and extragalactic sources. GALFACTS is a scientific pathfinder to the SKA in the area of cosmic magnetism. Key to magnetism science with the SKA is the technique of RM synthesis. The technique of RM synthesis is introduced and we discuss practical aspects of RM synthesis including efficient computational techniques and detection thresholds in the resulting Faraday spectrum. We illustrate the use of the technique by presenting the current development of the RM synthesis pipeline for GALFACTS and present early results.
\end{abstract}

\section{Introduction}
Linear polarization of radio sources contains information on magnetic
fields in these sources, and Faraday rotation of the plane of
polarization provides information on the direction and magnitude of
the magnetic field along the line of sight. As such, observations of
linear polarization of radio sources provide the most widely
applicable probe of cosmic magnetic fields on scales from galaxies to
clusters of galaxies. Finding polarized sources in survey images and
fitting their parameters forms the basis of this analysis. 

The Galactic ALFA Continuum Survey Consortium is using the Arecibo telescope and ALFA to carry out a
sensitive, high resolution, spectro-polarimetric survey of the region of the sky visible
with the Arecibo telescope \citep{Taylor_2010}.
A key observational objective of the GALFACTS is to image the polarized emission
from both discrete objects and the diffuse interstellar medium of our Galaxy and to
derive polarization properties, including Faraday Rotation Measures (RMs) for a vast population
of extragalactic sources. A detailed discussion on the GALFACTS project can be found in \cite{Taylor_2010}.

The objective of GALFACTS is to produce high sensitivity Stokes $I$, $Q$, $U$ and $V$ multi-frequency images
of the polarized emission over the full range of Galactic latitudes
from the mid-plane to the pole. GALFACTS has a bandwidth of $300$~MHz 
over 1225$--$1525~MHz, allowing imaging of Faraday rotation measures. 
The $1\sigma$ survey sensitivity over the full 
band is $\sim80~\mu$Jy per Stokes parameter ($\Delta T_{rms}\sim0.8$ mK) with a $3.4'$ resolution (maximum baseline 220m);
The total sky coverage of the full survey is 12,734 degrees which requires a total observing time 1600 hours. 
Since the GALFACTS survey produces terabyte sized data sets, the data processing
pipeline raises considerable challenges. Here we limit our discussion to the RM synthesis methods 
applied to the processed image data, a detailed description of the GALFACTS pipeline is given in \cite{Guram_2011}.

\section{Rotation Measure Synthesis}
Polarimetric radio observations enable the study of synchrotron emission radiated by
relativistic electrons as they are accelerated by magnetic fields. 
Multifrequency observations allow RM Synthesis \citep{burn_1966}  
to solve for the unknown Faraday depth, polarized intensity, and
polarization angle simultaneously. The source is typically identified
by the maximum value in the Faraday spectrum. 
Following \cite{brentjens_2005} the Faraday depth $\phi$ is defined as

\begin{equation}
\phi\left(\vec{r}\right) = 0.81 \int_{0}^{x} \vec{B_{||}} n_{e} \cdot \mathrm{d}\vec{r}\ \mbox{rad}\ \mbox{m}^{-2},
\end{equation}

where $\vec{B_{||}}$ is the line of sight magnetic field component, $n_e$ is the 
thermal electron density, $\mathrm{d}\vec{r}$ is an infinitesimal path length, 
with the integral taken from the observer to the point $x$.

The complex polarized intensity  $\mathcal{P}\left(\lambda^{2}\right) = Q + iU$ is the Fourier transform of the
Faraday dispersion function $F(\phi)$,

\begin{equation}
P\left(\lambda^{2}\right) = \int_{-\infty}^{\infty} F(\phi) e^{2 i  \phi \lambda^{2}} d\phi .
\end{equation}

The Faraday rotation measure $RM$ is defined as the slope of 
a polarization angle $\Psi$ versus $\lambda^{2}$:

\begin{equation}
RM\left(\lambda\right) = \frac{d \Psi}{d(\lambda^{2})} 
\end{equation}

where
\begin{equation}
\Psi = \frac{1}{2} \tan^{-1}\frac{U}{Q}.
\end{equation}

Once a polarized source has been detected, the observed polarized
intensity $p = \sqrt{Q^{2} + U^{2}}$ must be corrected for polarization
bias. Since $p$ is a positive-definite quantity, the noise in Stokes $Q$
and $U$ results in a positive value for $p$ even if no signal is
present. The statistics of $p$ with a signal $p_0$ and noise
$\sigma_{QU}$ is given by the Rice distribution
\citep{rice_1945}. This matter is complicated by the uncertainty in the Faraday depth of the source.
This introduces a stronger bias in the polarized intensity than the well known 
polarization bias but can be corrected for \citep{George_2011}. 

\section{Application to GALFACTS}
The first GALFACTS dataset (N1 region) is a cube of $5314$ by $1074$ pixels with $4096$ spectral channels. The large number of pixels means that to do effective RM synthesis is computationally expensive. To solve this RM synthesis can be run in parallel on each pixel in the image plane with the results being combined to form one Faraday depth cube. This is currently not part of the main GALFACTS pipeline \citep{Guram_2011} but will be integrated as an application on \url{http://www.cyberska.org} \citep{Kiddle_2011}. Another way to approach the RM synthesis is to average the Stokes $Q$ and $U$ data in frequency, correctly selecting the window to average so that no information is lost in the result Faraday depth cube. For the N1 region RM synthesis was completed over a range of $-1200$ to $1200$~rad.m$^{-2}$ with an averaging of $0.4$~MHz in the $Q$ and $U$. To visualize the Faraday depth cube is difficult with such a large dataset and to over come this limitation a moment map was created. A preliminary RM map from the N1 GALFACTS observations run is shown in Figure 1 with a typical compact source Faraday spectrum given in Figure 2. Residual imaging artefacts dominate the compact sources but lots of structure can be seen in the diffuse emission.

\begin{center}
	\includegraphics[angle=-90,width=1.2\linewidth]{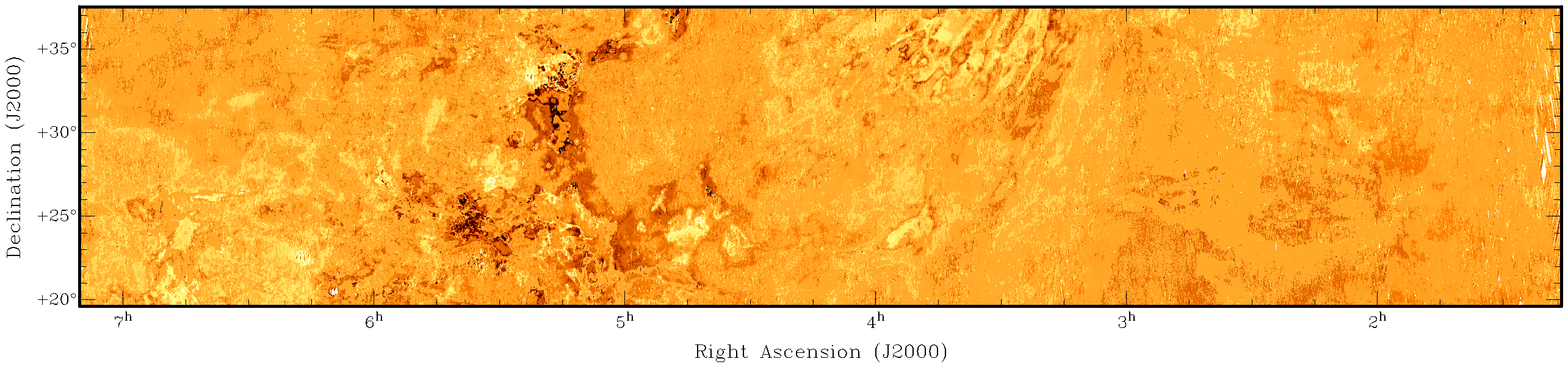} \\
{\bf Figure 1}: A moment map of a subset of the N1 region showing structure in the Faraday spectrum
\end{center}

\begin{center}
	\includegraphics[angle=0,width=0.5\linewidth]{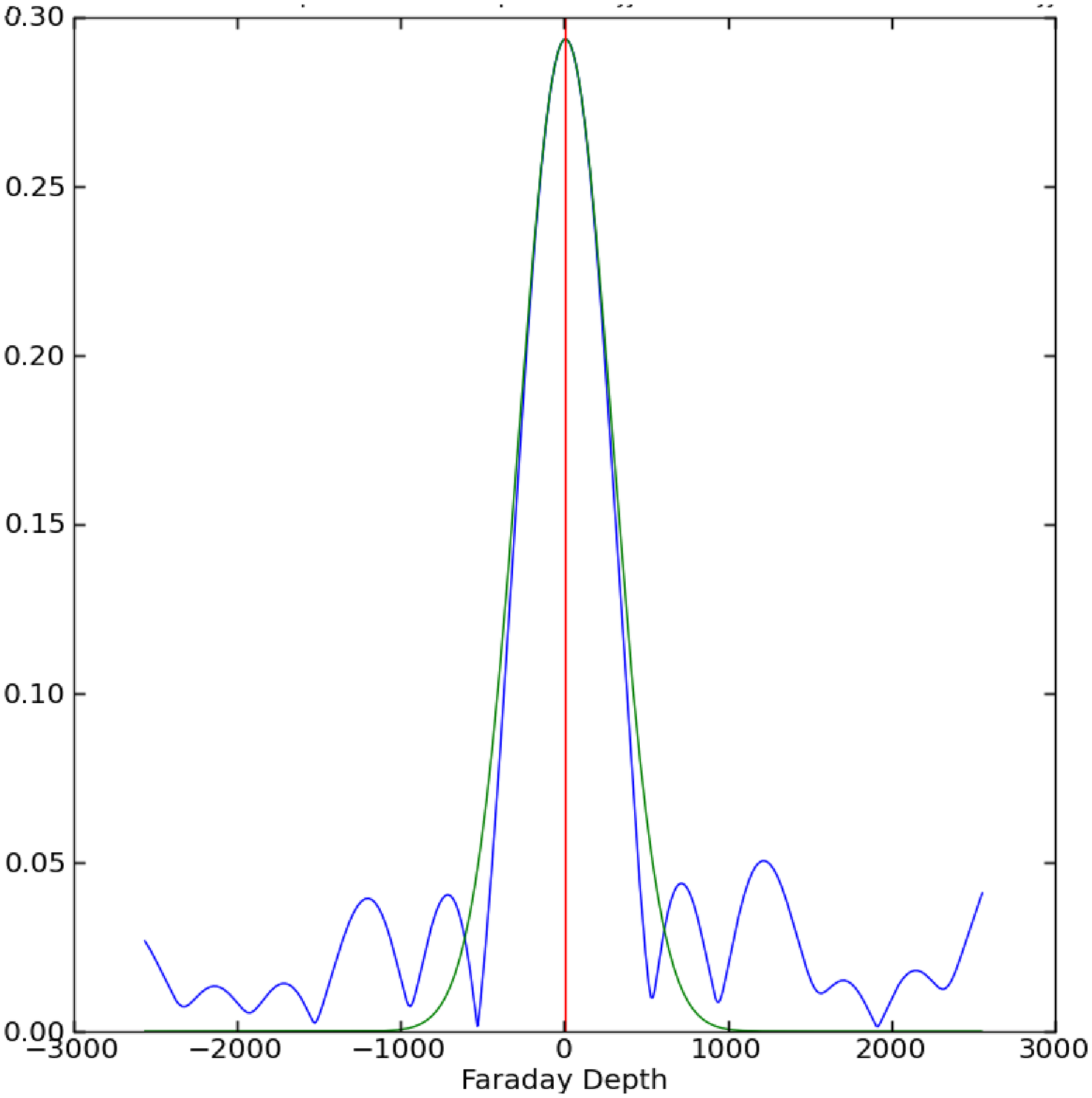}  \\
{\bf Figure 2}: The typical Faraday spectrum of a compact source in the preliminary N1 data
\end{center}

\bibliography{author}

\end{document}